\documentclass[aps,prl,twocolumn,groupedaddress,showpacs,floatfix,superscriptaddress]{revtex4-1}
\usepackage{epsfig}
\usepackage{amsmath,amssymb}
\usepackage{graphicx}
\usepackage[dvipsnames,usenames]{color}
\usepackage[normalem]{ulem}
\usepackage{soul} 
\emergencystretch=\maxdimen
\begin{document}

\title{Phonon dispersion and the competition between pairing 
and charge order}
\author{N.C. Costa} 
\email{natanael@if.ufrj.br}
\email{natanael.c.costa@gmail.com}
\affiliation{Instituto de F\'isica, Universidade Federal do Rio de Janeiro
Cx.P. 68.528, 21941-972 Rio de Janeiro RJ, Brazil}
\affiliation{Department of Physics, University of California, 
Davis, CA 95616,USA}
\author{T. Blommel} 
\affiliation{Department of Physics, University of California, 
Davis, CA 95616,USA}
\affiliation{Department of Physics, North Dakota State University, 
Fargo, ND 58105, USA}
\author{W.-T. Chiu}
\affiliation{Department of Physics, University of California, 
Davis, CA 95616,USA}
\author{G. Batrouni}
\affiliation{Universit\'e C\^ote d'Azur, INPHYNI, CNRS, 0600 Nice, France}
\affiliation{Beijing Computational Science Research Center, Beijing,
100193, China}
\author{R.T. Scalettar}
\affiliation{Department of Physics, University of California, 
Davis, CA 95616,USA}

\begin{abstract}
The Holstein Model (HM) describes the interaction between fermions and a
collection of local (dispersionless) phonon modes.  In the dilute limit,
the phonon degrees of freedom dress the fermions, giving rise to polaron
and bipolaron formation.  At higher densities, the phonons mediate
collective superconducting (SC) and charge density wave (CDW) phases.
Quantum Monte Carlo (QMC) simulations have considered both these limits,
but have not yet focused on the physics of more general phonon spectra.
Here we report QMC studies of the role of phonon dispersion on SC and
CDW order in such models.  We quantify the effect of finite phonon
bandwidth and curvature on the critical temperature $T_{\rm cdw}$ for
CDW order, and also uncover several novel features of diagonal long
range order in the phase diagram, including a competition between charge
patterns at momenta ${\bf q}=(\pi,\pi)$ and ${\bf q}=(0,\pi)$ which
lends insight into the relationship between Fermi surface nesting and the
wavevector at which charge order occurs.  We also
demonstrate SC order at half-filling in
situations where nonzero bandwidth sufficiently suppresses
 $T_{\rm cdw}$.  
\end{abstract}

\date{\today}

\maketitle


\noindent
\underbar{Introduction:}
Quantum Monte Carlo (QMC) methods have
evolved into a powerful tool to understand the physics of strongly
interacting quantum systems.  
Nevertheless, many qualitative questions remain largely unaddressed 
concerning electron-phonon models.
One of the most prominent concerns is the origin of charge-density
wave (CDW) formation,
especially in dimensions greater than one.
Increasingly attention has turned to alternatives to the original 
Peierls picture\cite{peierls55}.  Zhu {\it et al.} have 
proposed\cite{Zhu15,Zhu17} at least three classes of CDWs: (i) those
associated with the Peierls instability and  Fermi
Surface Nesting (FSN), typically in quasi-1D materials; 
(ii)  those driven by a momentum-dependent
electron-phonon coupling (EPC),
$g_{\mathbf{q}}$,
such as the quasi-2D material NbSe$_2$
\cite{Soumyanarayanan13,arguello14,weber11,Johannes08,Zhu15,johannes06,calandra09},
for which a CDW phase sets in
at $T_{\mathrm{cdw}}=33.5$\,K, even though ARPES measurements do
not show any sign of FSN\cite{Zhu15};
and (iii)  systems where electron
correlations are implicated in charge modulation, a primary example
being the cuprates\cite{dasilvaneto14}.
In addition to CDW physics, closely related current issues in
(high temperature) superconductivity (SC)
also invite a return to the study of electron-phonon interactions.
For instance, a momentum dependent EPC is
believed to be implicated in the dramatic
increase in the superconducting transition temperature $T_{sc}$ of FeSe
monolayers on SrTiO$_2$\cite{wang12,peng14,wang16}.

\begin{figure}[t]
\includegraphics[scale=0.25]{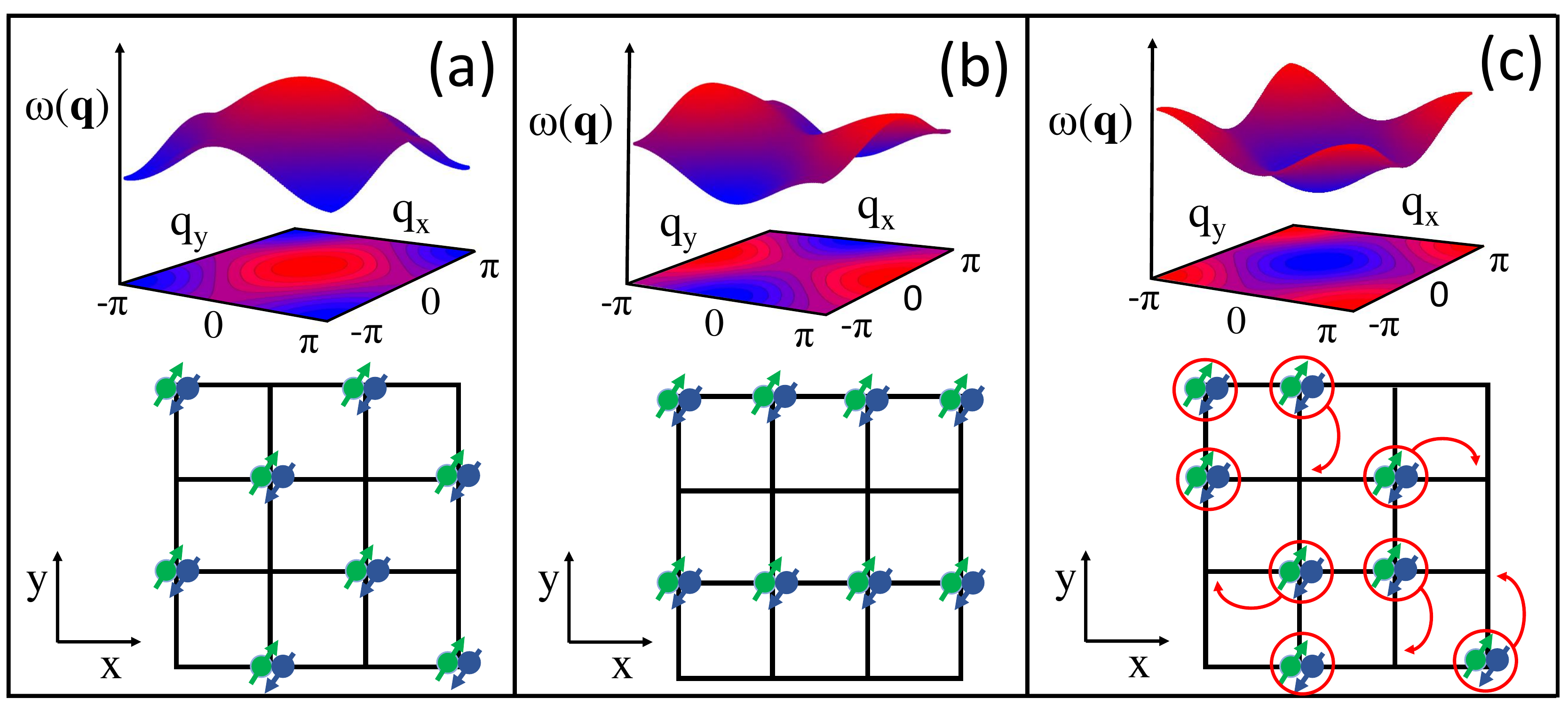} %
\caption{(Color online) Sketch of bare phonon dispersion (top)
and its resulting charge ordering (bottom) for
(a) downward curvature, (b) mixed curvature (saddle point at the origin), and
(c) upward curvature cases. The arrows on the latter
correspond to the (possible) hopping to any
available sites
and emphasize the possibility
of mobile pairs.}
\label{fig:phonon_sketch} 
\end{figure}

The Random Phase Approximation (RPA) criterion,
$4 g^{2}_{\mathbf{q}}/\omega(\mathbf{q}) 
 >  1 \,/\, \chi_{0}(\mathbf{q})$,
suggests that the shape of the bare phonon
dispersion, $\omega(\mathbf{q})$, should affect charge ordering,
and hence be important to the analysis of the second class of CDW above.
In view of this, here we explore a new scenario in which phonon dispersion
plays a primary role in determining the CDW ordering wavevector and
critical temperature, and where SC can supplant
diagonal long range order.
We extend QMC simulations\cite{Li16} of a 2D square 
lattice Holstein Model (HM)
to include phonon dispersion\cite{Raimbault95,Marchand13}.  
In the HM on a bipartite lattice, CDW
order dominates over SC at commensurate filling, similar to the
dominance of antiferromagnetism over pairing at
half-filling in the Hubbard Hamiltonian\cite{hirsch89,white89}.  
In that model, it is known\cite{lin87}
that off diagonal long range order (ODLRO) can be made more competitive
by adjusting the fermionic dispersion relation, e.g.~by introducing a
next-near-neighbor hopping $t^\prime$, or via doping.  Both these serve
to destroy the perfect nesting of the square lattice Fermi surface.
Here we adopt a different approach which is available in an
electron-phonon model -- tuning the phonon dispersion while retaining the
features of the bare electronic Fermi surface, i.e. its FSN.
The relevance of such approach can be infered by its effects on polaron
formation, as showed in a recent study,
Ref.\,\onlinecite{Marchand13}.
We examine in this Letter the many-electron problem,
with our results supporting the picture that the shape of
the phonon dispersion plays an important role in the CDW (or SC)
formation, i.e. being responsable for enhancing or suppressing it.

Fig.\,\ref{fig:phonon_sketch}
presents the qualitative pictures behind our key results:
Bare phonon dispersion with
(a) downward curvature in
going from ${\bf q}=(0,0)$ to ${\bf q}=(\pm\pi,\pm\pi)$ 
leads to an enhancement of
the CDW gap and increases $T_{\rm cdw}$ at half-filling;
(b) mixed curvature (saddle point at the origin),
i.e. upward in $\hat{x}$ and downward in $\hat{y}$ directions,
can lead to striped charge order --
further emphasizing that charge order and FSN
wavevectors do not have to be identical; and
(c) upward curvature, which suppresses the CDW gap and, 
for sufficiently large bandwidth,
can initiate a CDW-SC transition at {\it commensurate filling}.


\vskip0.03in \noindent
\underbar{Methodology:}
The Holstein model\cite{Holstein59},
\begin{align} \label{eq:Holst_hamil}
\nonumber \mathcal{H}_1 = &
-t \sum_{\langle \mathbf{i}, \mathbf{j} \rangle, \sigma} 
\big(d^{\dagger}_{\mathbf{i} \sigma} 
d^{\phantom{\dagger}}_{\mathbf{j} \sigma} + {\rm h.c.} \big)
- \mu \sum_{\mathbf{i}, \sigma} n_{\mathbf{i}, \sigma}
\\
&
+ \lambda \sum_{\mathbf{i}, \sigma} n_{\mathbf{i}, \sigma} \hat{X}_{\mathbf{i}}
+ \frac{1}{2} \sum_{ \mathbf{i} } \hat{P}^{2}_{\mathbf{i}}
+ \frac{\omega_{\, 1}^{2}}{2} \sum_{ \mathbf{i} } \hat{X}^{2}_{\mathbf{i}}
\,\,,
\end{align}
is one of the simplest tight binding descriptions of the
electron-phonon interaction.  A single electronic band, with fermionic
creation (destruction) operators at site ${\bf i}$,
$d^\dagger_{{\bf i},\sigma} \, (d^{\phantom{\dagger}}_{{\bf i},\sigma})
$, couples to independent oscillator degrees of freedom $\hat X_{\bf i},
\,\hat P_{\, \bf i}$.  We consider here a square lattice with periodic
boundary conditions, nearest neighbor (NN) electron hopping $t=1$
(to set the scale of energy), chemical potential $\mu$,
electron-phonon coupling $\lambda$, and local
phonon frequency $\omega_1$.

We generalize Eq.\,\eqref{eq:Holst_hamil} to ${\cal H}={\cal H}_1 + {\cal
H}_2$, to include a $\omega_2$ coupling between NN
displacements $\hat X_{\bf i}, \, \hat X_{\bf j}$, with  
\begin{align} \label{eq:Holst_hamil2}
\mathcal{H}_2 = 
 \frac{\omega_{\, 2}^{2}}{2} \sum_{ \langle \mathbf{i}, \mathbf{j} \rangle } 
\big( \hat{X}_{\mathbf{i}} \pm \hat{X}_{\mathbf{j}} \big)^{2} \, .
\end{align}
We will allow for both signs of this intersite term,
i.e. for
cases where the sign between neighboring sites
$\langle {\bf i,j} \rangle$ in the $\hat x$ and $\hat y$ directions
are equal or different.
Physically, the minus sign is the more natural one:  forces on 
atoms depend on their relative displacement.  On the other hand,
as we discuss below, the positive sign gives a mode with
a downward bending momentum 0 to $\pi$, the more typical
behavior for high frequency optical modes.

The inclusion of NN coupling $\omega_2 \neq 0$  leads to a finite phonon
bandwidth $\Delta\omega$.  In the absence of the electron-phonon
coupling, the quadratic bosonic Hamiltonian can be solved exactly, with
bare phonon dispersion relation
\begin{align}\label{eq:dispersion}
\omega({\mathbf{q}}) = \sqrt{ \omega^{\, 2}_{1} 
+ 2 \omega^{\, 2}_{2} \big[ 2 \pm \cos(q_{x}) \pm \cos(q_{y}) \big] }
\,\,.
\end{align}
Positive signs reduce $\omega(\pi,\pi)$,
making it energetically less costly to create a phonon
at the $M$ point, while negative signs favor 
modes at the zone center $\Gamma$ point, $\omega(0,0)$
[hereafter $\omega_{0}$].
A mixed sign breaks rotational symmetry, producing a phonon
in the $X$ (or $X^{\prime}$) point, 
$\omega(\pi,0)$ [or $\omega(0,\pi)$].  These three
cases, as depicted in Fig.\,\ref{fig:phonon_sketch},
are considered in this Letter.

\begin{figure}[t]
\includegraphics[scale=0.28]{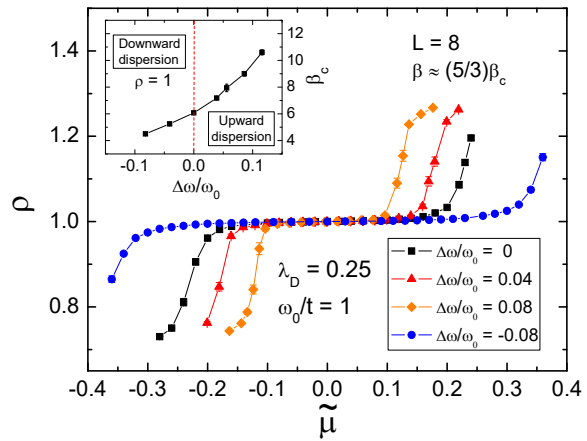} %
\caption{(Color online) Dependence of electronic density
$\rho$ on chemical potential $\tilde{\mu}$, fixing
$\lambda_{D}=0.25$, $\omega_{0}/t=1$ and
$\Delta\omega / \omega_{0}
= 0$ (black squares), 0.04 (red triangles),
0.08 (orange diamonds) and -0.08 (blue circles).
The energy scale is fixed for all cases, with
$\beta \approx \frac{5}{3} \beta_{c}$
(i.e. $\beta=10$, 12, 15 and 8, respectively).
Negative and positive signs for the bandwidth $\Delta \omega$
correspond to $\omega(\pi,\pi) < \omega_{0}$ and
$\omega(\pi,\pi) > \omega_{0}$, respectively.
Inset: Inverse critical temperature as a function of
$\Delta \omega$.
Here, and in all subsequent figures, when not
shown, error bars are smaller than symbol size.}
\label{fig:mu_var} 
\end{figure}

To facilitate discussion of the physics, we introduce
dimensionless parameters:
(i) the adiabaticity ratio $\omega_{0}/t$;
(ii) the phonon bandwidth $\Delta\omega / \omega_{0}$;
and (iii)
the dimensionless electron-phonon coupling,
\begin{align}\label{eq:EPIcoupling}
\lambda_{D}=\frac{1}{W} \frac{1}{N} \sum_{\mathbf{q}} 
\frac{\lambda^{2}}{\omega^{2}({\mathbf{q}})} \,\, ,
\end{align}
which is the polaron binding energy in units
of half electronic bandwidth.
Here $N=L^2$ is the number of sites, while the electronic bandwidth is $W=8t$.
One can show, through an
appropriate particle-hole transformation and shift of the phonon origin,
that a half-filled electronic band occurs at
$\mu=-\lambda^{2}/\omega_{0}^{2}$, for any dispersion relation
$\omega({\mathbf{q}})$.  We introduce
$\tilde \mu = \mu + \lambda^2/\omega_0^2$ so
that $\rho=1$ at $\tilde \mu=0$. In what follows, the Hamiltonian parameters
$\lambda$, $\omega_{1}$ and $\omega_{2}$ will be adjusted in order to
fix the dimensionless ratios, $\omega_{0}/t$, $\Delta\omega /
\omega_{0}$ and $\lambda_{D}$.

We examine the features of this generalization of the HM using
Determinant Quantum Monte Carlo (DQMC)
\cite{Blankenbecler81,Scalettar89,Noack91,Santos03}.
For details, see the Supplemental Material.
The nature of charge ordering is investigated by the
equal time charge-density correlation function,
\begin{align} \label{eq:CDW_cor}
S(\mathbf{q})= \frac{1}{N} 
\sum_{\mathbf{i},\mathbf{j}} 
e^{i \mathbf{q} \cdot (\mathbf{i} - \mathbf{j})} \,
\langle \, n_{\mathbf{i}} \, n_{\mathbf{j}} \, \rangle,
\end{align}
while pairing features are analysed by the $s$-wave
superconducting pair susceptibility
\begin{align} \label{eq:SC_sus}
P_{s} = \frac{1}{N} 
\int^{\beta}_{0} \mathrm{d}\tau \,
\langle \Delta(\tau) \Delta^{\dagger}(0) + \mathrm{H.c.} \rangle,
\end{align}
with
$ \Delta(\tau) = \sum_{\mathbf{i}}
c^{\phantom{\dagger}}_{\mathbf{i}\downarrow}(\tau) 
c^{\phantom{\dagger}}_{\mathbf{i}\uparrow}(\tau)$.


Before presenting our main results on the effects of phonon dispersion
on charge and pairing order, we revisit the dispersionless
($\omega_2=0$) HM.  Quite remarkably, it is only very recently that
early simulations\cite{Scalettar89,Noack91,Vekic92,Niyaz93}, for which
values of the inverse critical temperature 
$\beta_c$ differed by almost 20\%, have been
followed up to obtain more accurate results for the critical
temperature\cite{Costa17,weber17}.
Fixing $\lambda_{D}=0.25$, $\omega_{0}/t=1$, at
half-filling ($\rho=1$), we find here that
$\beta_{c}=6.0 \pm 0.1$.  (See Supplemental Material.)
This value of $\beta_c$ is somewhat lower than the earliest DQMC
results\cite{Noack91,Vekic92},
but in agreement with more recent simulations\cite{weber17,Costa17},
and will be used as a benchmark when analysing the effects of
phonon dispersion.
The higher accuracy follows from both advances in raw computer speed,
but also improved understanding of the nature of the global moves
required to reduce autocorrelation times.

\begin{figure}[t]
\includegraphics[scale=0.28]{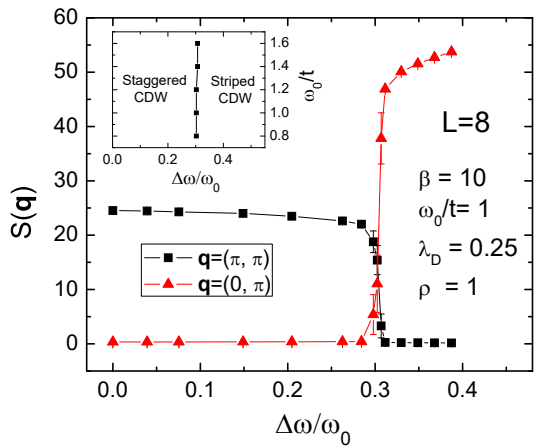} %
\caption{(Color online) CDW structure factor as a function of phonon
bandwidth for the mixed curvature dispersion case,
i.e. upward in $\hat{x}$ and
downward in $\hat{y}$ directions.
A phase transition from staggered
to striped order occurs at around $\Delta\omega/\omega_{0} =0.30$,
independent of $\omega_0/t$ (inset).}
\label{fig:stripe1} 
\end{figure}


\vskip0.03in \noindent
\underbar{Effect of Dispersion on Charge Correlations:}
We first consider the case where Eq.\,\eqref{eq:Holst_hamil2}
has the same sign for both spatial coordinate directions. 
A positive coupling ($\hat X_{\bf i} + \hat X_{\bf j}$) in
Eq.\,\eqref{eq:Holst_hamil2} corresponds to
$\omega(\pi,\pi) < \omega_{0}$ and is
expected to enhance CDW order.
On the other hand, a negative coupling leads to
$\omega(\pi,\pi) > \omega_{0}$ and charge order at the $M$ point.
These two cases correspond to Fig.\,\ref{fig:phonon_sketch}\,(a)
and (c), respectively.
We define
$\Delta \omega = \omega(\pi,\pi) - \omega_{0}$,
i.e.~$\Delta \omega > 0$ ($<0$) for upward (downward) phonon dispersion.
The effect of $\omega_2 \neq 0$ is quantified in
Fig.\,\ref{fig:mu_var}, which shows the charge gap induced in
$\rho(\tilde{\mu})$ by the electron-phonon coupling
\footnote{These results also agree with Max. Entropy Method, as presented
in the Supplemental Material.}.
This CDW gap grows or shrinks depending on the shape of the phonon dispersion,
i.e. if it is downward or upward, respectively.
As presented in the inset of Fig.\,\ref{fig:mu_var},
this behavior is accompanied by changes in
$\beta_c = 1/T_{\rm cdw}$, obtained by the same data collapse
as used for the dispersionless case.  See Supplemental Material.
It is remarkable that $T_{\rm cdw}$ can decrease by a factor of two
with a relatively small $\Delta \omega / \omega_{0} \approx 0.1$.


\begin{figure}[t]
\includegraphics[scale=0.28]{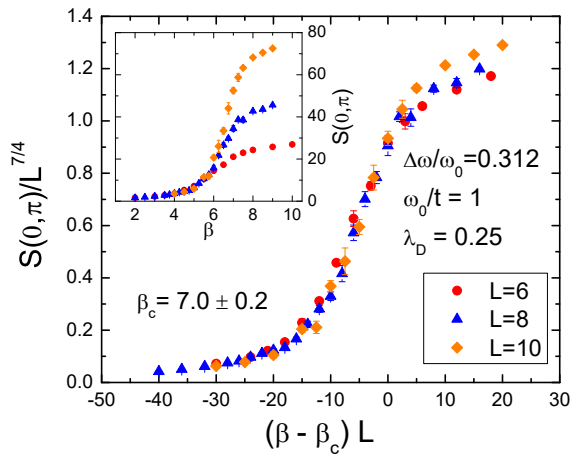} %
\caption{(Color online) Data collapse of the DQMC results of
$S(0,\pi)$ for the mixed curvature (saddle point) case, fixing
the 2D Ising critical exponents.
Inset: Charge structure factor as function of $\beta$.
Here $\Delta\omega/\omega_{0} = 0.312$.}
\label{fig:stripes3} 
\end{figure}

A mixed sign, in which phonon dispersion terms in
Eq.~\eqref{eq:Holst_hamil2} take the form $\hat X_{\bf i} - \hat X_{\bf
j}$ for ${\bf j} = {\bf i} + \hat x$ and $\hat X_{\bf i} + \hat X_{\bf
j}$ for ${\bf j} = {\bf i} + \hat y$, results in a phonon spectrum with
a saddle point at $\mathbf{q}=(0,0)$, with mimima at $\mathbf{q}=(0,\pm
\pi)$ and maxima at $\mathbf{q}=(\pm \pi,0)$,
see e.g. Fig.\,\ref{fig:phonon_sketch}\,(b).
Figure \ref{fig:stripe1}
shows the charge structure factors for checkerboard [${\bf
q}=(\pi,\pi$)] and striped [${\bf q}=(0,\pi$)], order as a function of
the phonon bandwidth $\Delta\omega = \omega(\pi,0) - \omega(0,\pi)$, for
fixed $\lambda_{D}=0.25$, $\omega_{0}/t =1$, $\beta=10$ and $L=8$.  In contrast
to the case of identical sign, where small $\Delta \omega/\omega_{0}
\sim 0.1$ had a large effect on the gap and $T_{\rm cdw}$, the charge
correlations here are initially almost independent of $\Delta\omega$ up
to $\Delta\omega/\omega_{0} \sim 0.25$.  However, at $\Delta\omega /
\omega_{0} \sim 0.30$ a strong suppression of $S(\pi,\pi)$ occurs, with
a corresponding rapid rise in $S(0,\pi)$.  This transition point is
almost independent of $\omega_{0}/t$, as displayed in Fig.\,\ref{fig:stripe1} (inset).
One should notice that the bare fermion dispersion relation is of course {\it independent} of
$\Delta\omega$, i.e.~it retains a nesting at $(\pi,\pi)$ and a van-Hove
singularity at $\rho=1$.  The onset of striped charge order is
initiated by changes in the phonon dispersion, not any alteration to
FSN.

We can also obtain the transition temperature
for the striped phase.
The inset of Fig.\,\ref{fig:stripes3} shows raw data for
$S(0,\pi)$ on different lattice sizes as a function of $\beta$, for
$\Delta\omega/\omega_{0} = 0.312$, slightly after 
entry into the striped phase. The corresponding
scaling (data collapse) is presented in Fig.\,\ref{fig:stripes3},
using the same Ising exponents
as for the checkerboard case, Eq.\,(S3),
indicating a finite temperature phase
transition at $\beta_{c} \approx 7.0$.

\begin{figure}[t]
\includegraphics[scale=0.28]{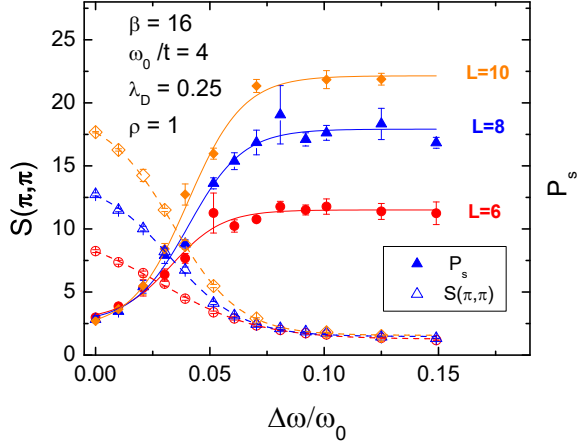}
\caption{(Color online) 
The CDW-SC transition at half-filling:
As $\Delta \omega/\omega_0$ increases, $S(\pi,\pi)$ is strongly suppressed
while pairing susceptibility $P_s$ is enhanced.
Both quantities are in
the same scale.
Circles, triangles and diamonds
correspond to $L=$6, 8 and 10, respectively.
The filled (open) symbols represent
$P_s$ [$S(\pi,\pi)$].
The lines are just guide to the eye.
} 
\label{fig:SC3} 
\end{figure}

{\it This striped phase,
with $\mathbf{q}_{\rm cdw}=(0, \pi) \neq 2 \, \mathbf{k}_{F}$,
provides an explicit and quantitative
illustration of a non-Peierls CDW instability.}
Recent experiments have exposed a similar behavior in a
variety of materials, i.e.~a charge order arising away from $2 \, \mathbf{k}_{F}$
and whose origin can not be related to the FSN, such as in the quasi-2D
materials NbSe$_2$, CeTe$_3$, Cr, and U, 
and also in one dimensional model systems like Au/Ge(001)
\cite{Johannes08,Zhu15, Maschek15,Lamago10,Raymond11,Blumenstein11}.
In particular, NbSe$_2$ does not exhibit FSN or any divergence in the electronic
susceptibility\cite{johannes06}, nor a metal-insulator transition.
Nevertheless,
CDW order sets in at $T_{\mathrm{cdw}}=33.5$\,K.
The appearance of this phase, outside the usual Peierls paradigm,
is then instead ascribed 
to strong EPC\cite{Zhu15,Zhu17}.
As noted earlier, the RPA criterion for CDW order
suggests an intimate connection between momentum dependent $g_{\bf q}$ 
and phonon dispersion $\omega({\bf q})$, so that the results
of Fig.~4 provide a confirmation that additional momentum structure
plays a crucial role in the CDW ordering wave vector.
We discuss, in the Supplemental Material,
possible differences between $g_{\bf q}$ and $\omega({\bf q})$,
which lend some additional complexity.

\begin{figure}[t]
\includegraphics[scale=0.28]{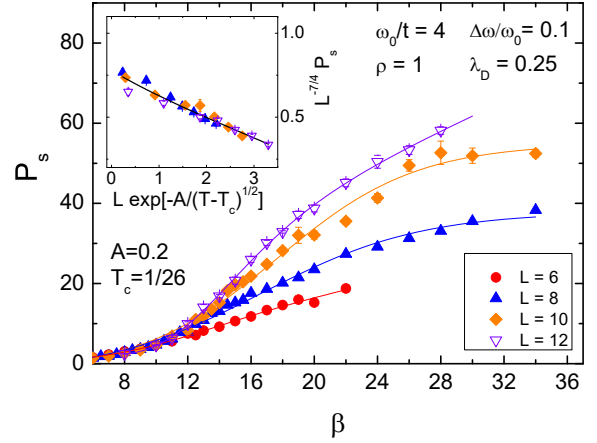}
\caption{(Color online) $s$-wave pair susceptibility as function of the
inverse of temperature for $\Delta\omega/\omega_{0}=0.1$, $\omega_{0}/t=4$
and $\lambda_{D}=0.25$. Inset: the data collapse of the raw DQMC results
by Kosterlitz-Thouless scaling for $\beta \geq 16$ (and $L=$8, 10 and 12).
The full lines are just guide to the eye.}
\label{fig:SC2} 
\end{figure}


\vskip0.03in \noindent
\underbar{Effect of Dispersion on Pairing:}
We now turn to SC order.  As noted earlier, it is uncommon for ODLRO
to appear in fermionic models at half-filling
in bipartite geometries like the square lattice which instead
favor diagonal order.  Nevertheless, the data of 
Fig.\,\ref{fig:mu_var} show a rise in $\beta_c$ with the increased
energetic cost
for $(\pi,\pi)$ CDW formation from the upward phonon dispersion. 
A natural question is whether that cost 
eventually becomes prohibitive, opening the door to SC. 

To address this, we increase the SC scale of energy and consider
phonon frequency $\omega_{0}/t=4$.
For this case, we obtain $T_{\rm cdw}\sim 1/13$ for the
dispersionless HM ($\omega_2=0$), without SC;
see Supplemental Material.
However, for the dispersive case, $S(\pi,\pi)$
is strongly suppressed at $\Delta \omega/\omega_0 \gtrsim 0.05$,
while $P_s$ is enhanced and grows with lattice size, as
displayed in Fig.\,\ref{fig:SC3}, for fixed $\beta=16$.
In words, a CDW-SC transition should occur when $\Delta \omega/\omega_0$
increases.
As presented in Fig.\,\ref{fig:SC2},
at $\Delta \omega/\omega_0 = 0.10$, for instance,
$P_{s}$ grows with lattice size
for $\beta \gtrsim 12$.
In order to establish quasi-long-range order for
this case,
the appropriate scaling analysis is a Kosterlitz-Thouless (KT) behavior,
\begin{align}\label{eq:KT_scaling}
P_{s} = L^{2-\eta} f\big(L / \xi),
\end{align}
with $\eta=1/4$ and
\begin{align}\label{eq:KT_xi}
\xi \sim 
\mathrm{exp}\bigg[ \frac{A}{(T - T_{c})^{1/2}} \bigg],
~~~~ T \to T^{+}_{c}.
\end{align}
The inset of Fig.\,\ref{fig:SC2} displays the KT scaling of the
$P_s$ raw data for $\beta \geq 16$.
Here, the parameters $A=0.2$ and $T_{c}=1/26$ yield
the best data collapse.  
This result provides strong evidence for 
the onset of SC at half-filling in the HM, when phonon
dispersion is taken into account;
see also the Supplemental Material.
We should mention that recent results\cite{Lin16,Lin17,Mendl17,Esterlis17,Ohgoe17,Karakuzu17}
have also examined the onset of SC in HM, but they have not consider the effects of
phonon dispersion.


\vskip0.03in \noindent
\underbar{Conclusions:}
This paper has provided a significant extension of QMC simulations
of electron-phonon Hamiltonians by evaluating the effects
of phonon dispersion on charge and pairing order in the Holstein
model.  The results offer several interesting features, including
a CDW-SC transition at half-filling and
transitions between CDW phases at different ordering momenta,
which are controlled by the bare phonon dispersion.
Our findings of non-Peierls CDW phase, despite of existence of FSN
on the bare electron dispersion, is of particular interest given 
recent work calling into question the traditional view of CDW formation
\cite{Johannes08,Zhu15,Maschek15,Lamago10,Raymond11,Blumenstein11}.
In view of this, our results present further insight into the (complex)
nature of CDW formation, exibithing a new avenue to understand and,
ultimately, control it.


\vskip0.03in \noindent
\underbar{Acknowledgements:}
We thank E. da Silva Neto for useful
suggestions concerning the manuscript.
RTS was supported by Department of Energy grant 
DE-SC0014671, and NCC by the Brazilian agencies Faperj and CNPq.
GGB acknowledges support from the University of the C\^ote d'Azur IDEX Jedi
and Beijing CSRC.
RTS and GGB acknowledge useful input from K.D.\,Lewis.


\bibliography{bibCosta_Hols}

\clearpage
\noindent
{\Large Supplemental Material for: Phonon dispersion and the competition between pairing and charge order}

\makeatletter
\renewcommand{\thefigure}{S\@arabic\c@figure}
\renewcommand{\theequation}{S\@arabic\c@equation}
\setcounter{equation}{0}
\setcounter{figure}{0}
\makeatletter

\begin{figure}[h]
\includegraphics[scale=0.18]{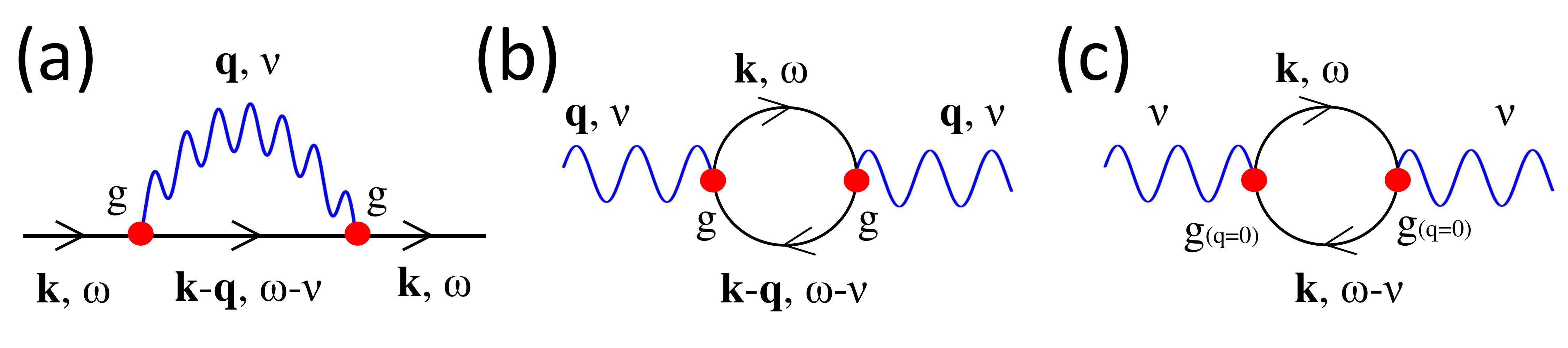} %
\caption{(Color online) 
(a) Electron self-energy to second order in
perturbation theory, as well as
phonon self-energies for (b) phonon dispersion and (c) momentum-dependent
electron-phonon coupling cases.}
\label{fig:sigma} 
\end{figure}

\noindent
\underbar{Phonon dispersion and electron-phonon coupling:}
Including momentum dependence in the phonon dispersion relation has 
some qualitative connections to a momentum-dependent electron-phonon coupling,
but is rather different when considered in detail.  This can be seen,
for instance, through a calculation of the lowest-order
electron self-energy $\Sigma(k,\omega)$ in Fig.\,\ref{fig:sigma}\,(a),
\begin{eqnarray}
\Sigma^g(k,\omega) &\sim \int \, dq \, d\nu \, |g(q)|^2 
\, \frac{1}{\omega-\nu- \epsilon_{k-q}}
\, \frac{2 \, \omega_0}{\nu^2 - \omega_0^2}
\nonumber \\
\Sigma^{\,\omega}(k,\omega) &\sim \int \, dq \, d\nu \, |g|^2 
\, \frac{1}{\omega-\nu- \epsilon_{k-q}}
\, \frac{2 \, \omega(q)}{\nu^2 - \omega(q)^2} ,
\end{eqnarray}
where $\Sigma^g$ and
$\Sigma^{\, \omega}$ are the forms for $g(q)$ and $\omega(q)$
respectively.
In the $\nu \rightarrow 0$ limit, where the phonon
carries no energy, we have that
$\Sigma^g = \Sigma^{\, \omega}$ under the condition
$|g(q)|^2 / \omega_0 = |g|^2 / \omega(q)$.
However it is evident 
that for nonzero $\nu$ the two self-energies are not equal.

Such difference is even more accentuated in the phonon self-energy.
For phonon dispersion case, Fig.\,\ref{fig:sigma}\,(b), the phonon
self-energy is
\begin{align}
\Pi^{\,\omega}(q,\nu) = \int \, dk \, d\omega \, g^2 
\, \frac{1}{\omega - \epsilon_{k}}
\, \frac{1}{\omega - \nu_{q} - \epsilon_{k-q}},
\end{align}
while for the momentum-dependent electron-phonon coupling case,
Fig.\,\ref{fig:sigma}\,(c), one obtains
\begin{align}
\Pi^{g}(\nu) = \int \, dk \, d\omega \, |g(q=0)|^2
\, \frac{1}{\omega - \epsilon_{k}}
\, \frac{1}{\omega - \nu - \epsilon_{k}}.
\end{align}

We have written these expressions in conventional real time
notation.  The expressions appropriate for DQMC are the 
imaginary-time (Masubara frequency) analogs, but the 
point, that $\Sigma^g$ and $\Sigma^{\,\omega}$ 
as well as $\Pi^{\,\omega}$ and $\Pi^{g}$
are in general inequivalent, is the same.
At higher order, e.g.~in vertex corrections,
the differences between $\omega(q)$ and
$g(q)$ approaches will become even more complex.

\begin{figure}[t]
\includegraphics[scale=0.28]{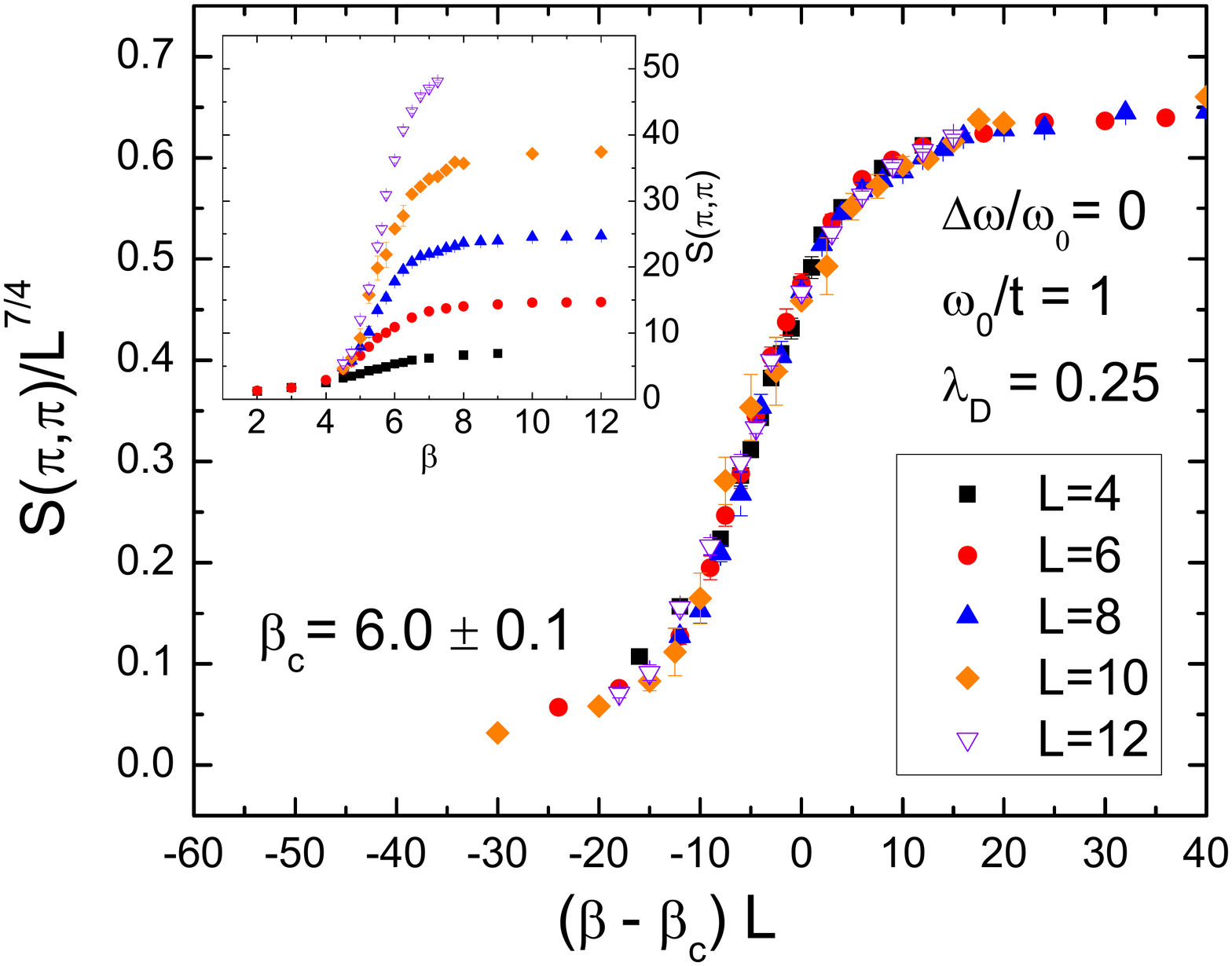} %
\caption{(Color online) Data collapse of the charge-density correlation
function of the dispersionless Holstein model at half filling and
fixing $\lambda_{D}=0.25$ and $\omega_{0}/t=1$ within the 2D Ising universality class. 
Inset: $S(\pi,\pi)$ as function of inverse of temperature.
Here, and in all subsequent figures, when not shown, error bars are smaller
than symbol size.
}
\label{fig:CDW_dispersionless} 
\end{figure}
\noindent
\underbar{Determinant Quantum Monte Carlo:}
We examine the features of the Holstein model (HM)
using Determinant Quantum Monte Carlo (DQMC)
simulations\cite{Blankenbecler81,Scalettar89,Noack91,Santos03}.  We first perform
the usual mapping\cite{creutz81} of the quantum oscillator degrees of
freedom onto a path integral in imaginary time by discretizing the
inverse temperature $\beta=\Delta\tau L_\tau$.  The degrees of freedom
of the fermions moving in this fluctuating space and imaginary time
phonon field can be integrated out analytically, so that the partition
function is
\begin{align}\label{eq:partition_func_Hols}
{\cal Z}  = &\int \mathrm{d}\{x_{{\bf i},l}\} \,e^{-\Delta \tau S_{B}} \,
\bigg[ \mathrm{det}\big(I + B_{1} B_{2} \cdots B_{L} \big) \bigg]^2,
\nonumber \\
S_{B} = &\sum_{{\bf i}}
 \sum_{l=1}^{L_\tau} \bigg[ 
\frac{1}{2} \bigg( \frac{x_{{\bf i},l} - x_{{\bf i},l+1}}{\Delta \tau} \bigg)^2 
+ \frac{\omega_{1}^{2}}{2} x^{2}_{{\bf i},l}  \bigg]
\nonumber \\
+ 
&\sum_{\langle {\bf i,j}\rangle}
\sum_{l=1}^{L_\tau} \bigg[ 
\frac{\omega_{2}^2}{2} \big(x_{{\bf i},l} \pm x_{{\bf j},l} \big)^2 
\bigg].
\end{align}
Here $\int \mathrm{d}\{x_{i,l}\}$ is the integral over the
(continuous) space- and imaginary time-dependent phonon field.
$S_B$ is the bosonic action.  The determinant appears as a square
because the two fermionic species $\sigma=\uparrow,\downarrow$
experience the same phonon field and contribute identically to ${\cal
Z}$.  An important consequence is the {\it absence of a sign problem at
any filling}.  The matrices $B_{l}$ are each a product of an exponential
of the hopping term in Eq.\,(1) and an exponential of a
site-diagonal matrix containing the phonon variables 
at that imaginary time slice $l$.

The physical quantities of interest are obtained by sampling the phonon
fields $\{x_{{\bf i},l}\}$ and measuring combinations of the
equal time fermion Green's function $G = \big[ I + B_1 B_2 \cdots
B_{L_\tau} \big]^{-1}$.  We keep $\Delta \tau$ small enough
(in most of cases $\Delta \tau = 0.1$) so that
Trotter errors are less than the statistical errors from the Monte Carlo sampling.
We should stress that, in addition to local moves of single phonon coordinates,
we also implemented {\it global} moves\cite{Scalettar91}, which reduce autocorrelation times.
Despite of that, our simulations were performed with a large number of
Monte Carlo steps, in most of cases larger than $2\times 10^5$.

\begin{figure}[t]
\includegraphics[scale=0.32]{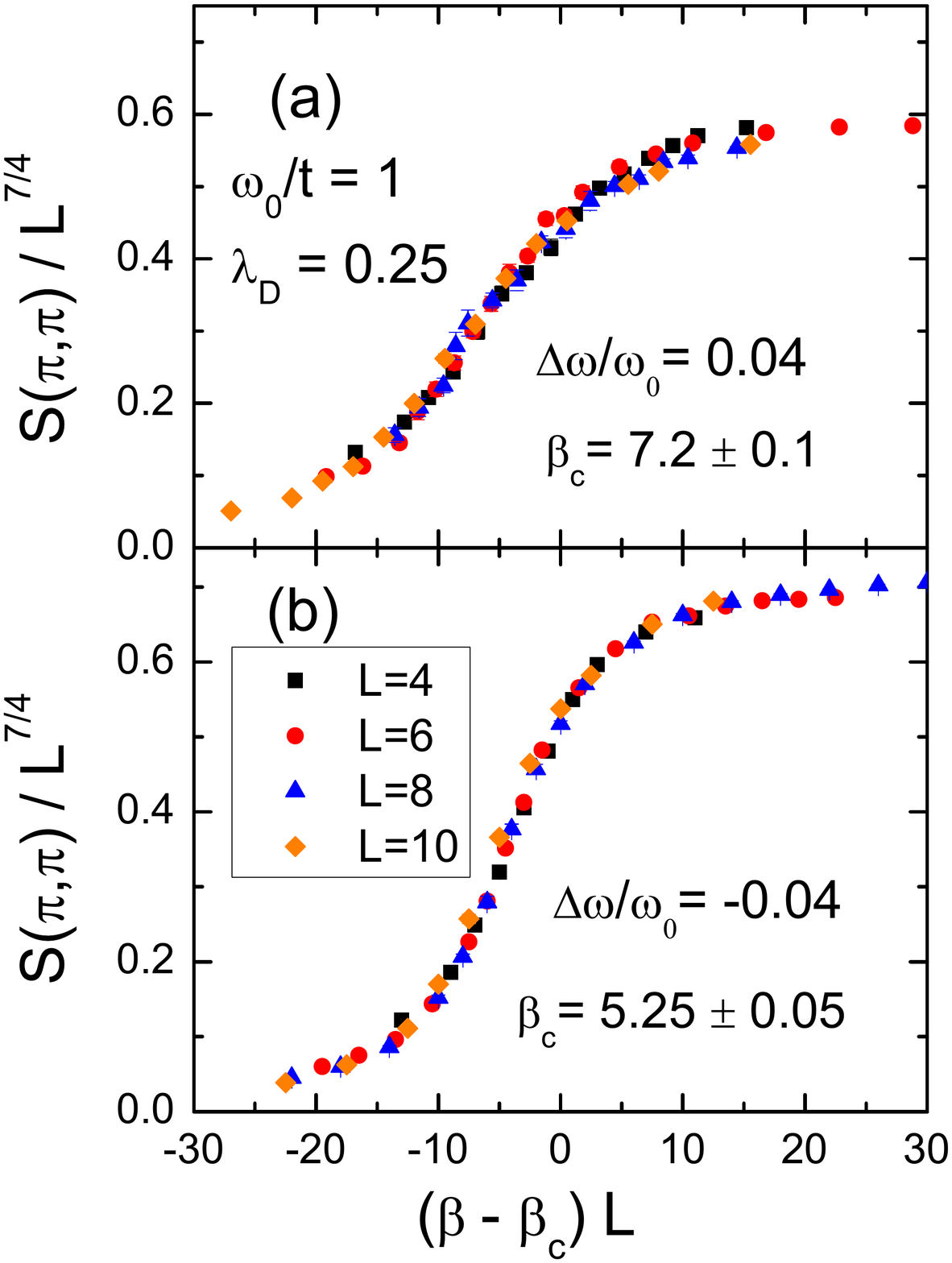} %
\caption{(Color online) Data collapse of $ S(\pi,\pi) $ for different
lattice sizes at (a) $\Delta\omega / \omega_{0} = 0.04$ and
(b) -0.04, and fixing $\lambda_{D}=\omega_{0}/t=1$.}
\label{fig:dispersion1} 
\end{figure}

\vskip0.03in \noindent
\underbar{Data collapse ($\Delta\omega/\omega_{0} = 0$):}
The inset of Fig.\,\ref{fig:CDW_dispersionless} shows 
the charge-density correlation function 
for the parameters of Refs.\,\onlinecite{Noack91}
and \onlinecite{Vekic92}, $\lambda_{D}=0.25$ and $\omega_{0}/t=1$ and
half-filling, $\rho=1$.
$S(\pi,\pi)$ exhibits a strong dependence on spatial lattice size
for $\beta \gtrsim 5$. 
A precise determination of the transition temperature is obtained via
finite size scaling.
Since the HM exhibits a finite temperature phase transition which breaks
$\mathbb{Z}_{2}$ (spin-inversion) symmetry, the transition belongs to the 
two-dimensional Ising universality class, and hence,
\begin{align} \label{eq:scaling}
S(\pi,\pi) = L^{2-\eta} f(L(\beta - \beta_{c})^{\nu}),
\end{align}
with $\eta=1/4$ and $\nu = 1$.
Fixing these exponents, the data collapse
is presented in Fig.\,\ref{fig:CDW_dispersionless},
with $\beta_{c}=6.0 \pm 0.1$.

\vskip0.03in \noindent
\underbar{Data collapse ($\Delta\omega/\omega_{0} \neq 0$):}
Regardless the phonon dispersion, i.e. if it is downward or upward, the
the staggered CDW phase still has a discrete order parameter and breaks
the $\mathbb{Z}_{2}$ (spin-inversion) symmetry. Thus, the $T_{\mathrm{cdw}}$ can be obtained by
performing a data collapse of the DQMC data points using the scaling function
of Eq.\,\eqref{eq:scaling}, similarly to the previous case.
For instance, Fig.\,\ref{fig:dispersion1} exhibits the data collapse for
(a) $\Delta\omega/\omega_{0} = 0.04$ [downward case] and
(b) $\Delta\omega/\omega_{0} = -0.04$ [upward case]
at fixed $\lambda_{D}=0.25$ and $\omega_{0}/t=1$.
In the former it is obtained $\beta_{c} \approx 7.2$, while in the
latter it is $\beta_{c} \approx 5.25$.
The same procedure is performed to obtain $T_{\mathrm{cdw}}$
for other values of $\Delta\omega/\omega_{0}$.


\begin{figure}[t]
\includegraphics[scale=0.28]{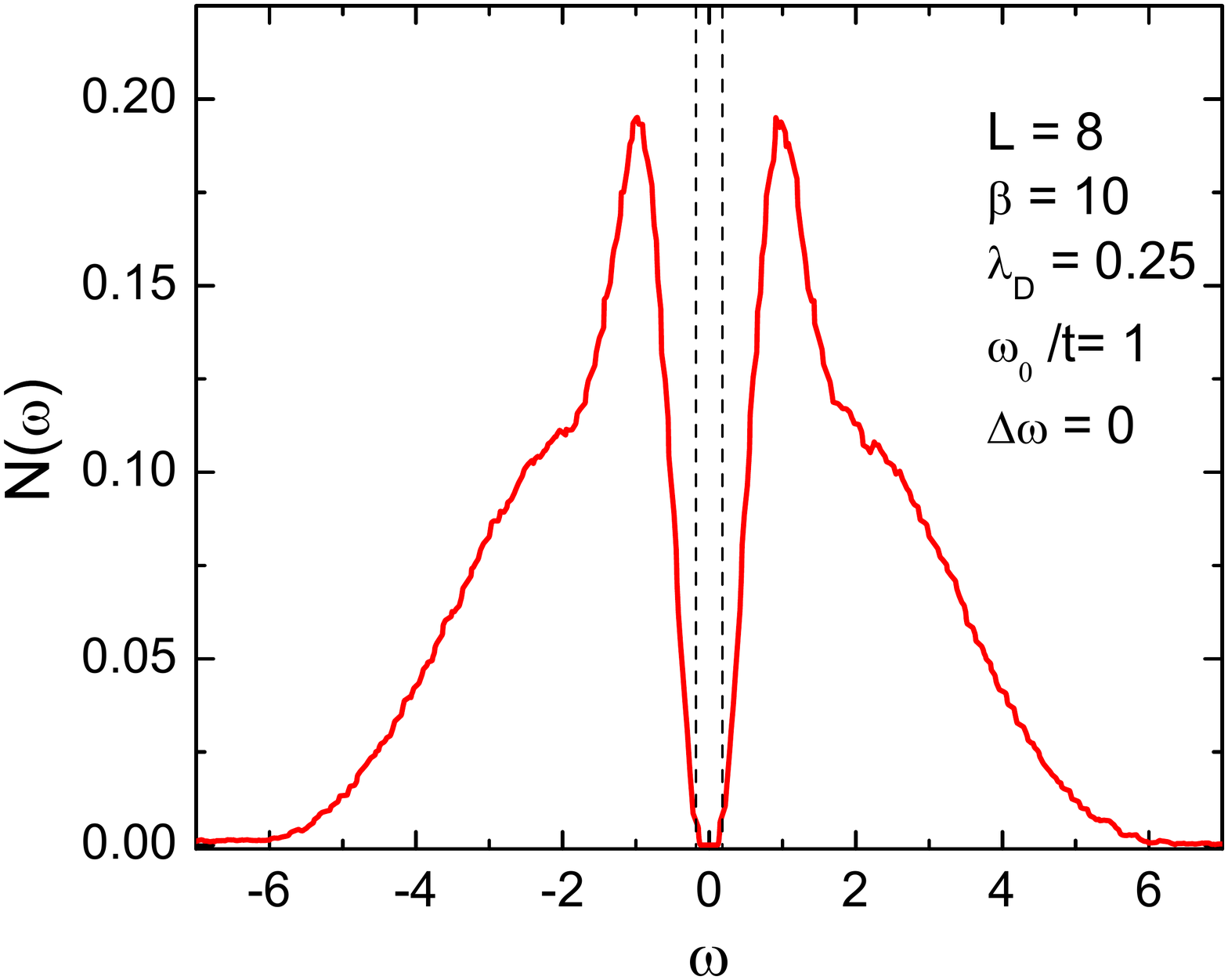} %
\caption{(Color online) Density of states of the dispersionless Holstein
model at $\beta=8$ and fixed $\lambda_{D}=0.25$ and $\omega_{0}/t=1$.
The vertical dashed lines enclose the charge gap.}
\label{fig:DOS}
\end{figure}

\begin{figure}[t]
\includegraphics[scale=0.34]{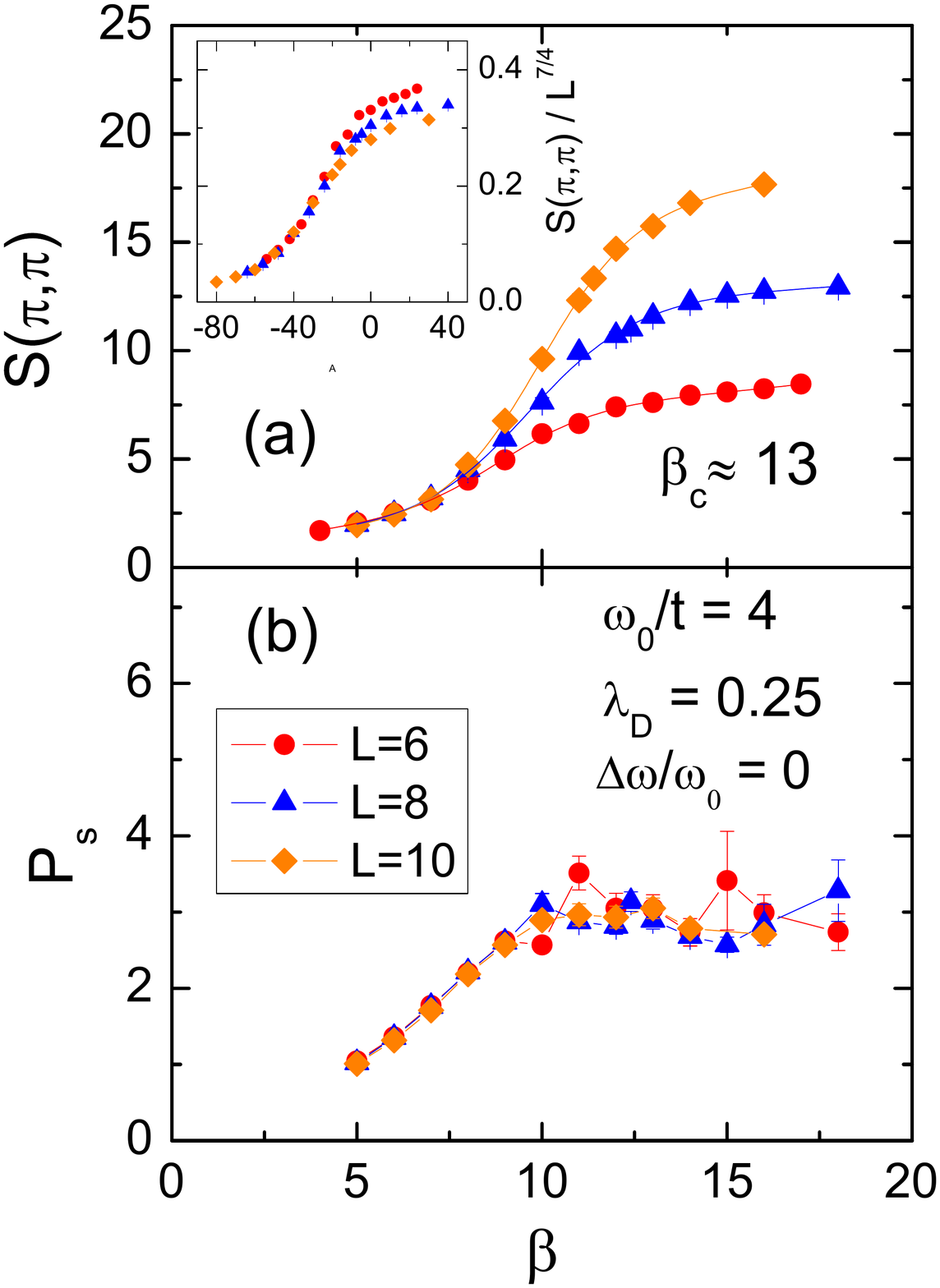} %
\caption{(Color online) (a) Staggered chage-density structure factor
behavior of the dispersionless Holstein model at different lattice sizes
and fixed $\omega_{0}=4$ and $\lambda_{D}=0.25$. Inset: Its corresponding
scaling (data collapse) using Eq.\,\eqref{eq:scaling}. (b) The $s$-wave pair susceptibility
for the same parameters of panel (a). The solid lines are just guides to the eye.}
\label{fig:wo40wa00}
\end{figure}

\vskip0.03in \noindent
\underbar{Density of States:}
It is also worth examining the spectral properties of the system, namely
its density of states (DOS), as an independent check to the results of
Fig.\,2. To this end, we need to perform an analytic continuation of
the imaginary-time dependent Green's function by inverting the integral equation
\begin{align}
G(\tau) = \int \mathrm{d}\omega\,
N(\omega) \,
\frac{e^{-\omega \tau}}{e^{\beta \omega} + 1}.
\label{eq:aw}
\end{align}
This task can be performed by
using the Maximum Entropy Method (MEM)\cite{Jarrell96}.
For instance, Fig.\,\ref{fig:DOS} exhibits the DOS of the dispersionless
Holstein model at $\beta=8$ and fixed $\lambda_{D}=0.25$ and $ \omega_{0}/t=1$.
The resulting charge gap using MEM is in good agreement with the one
presented in Fig.\,2.


\begin{figure}[t]
\includegraphics[scale=0.30]{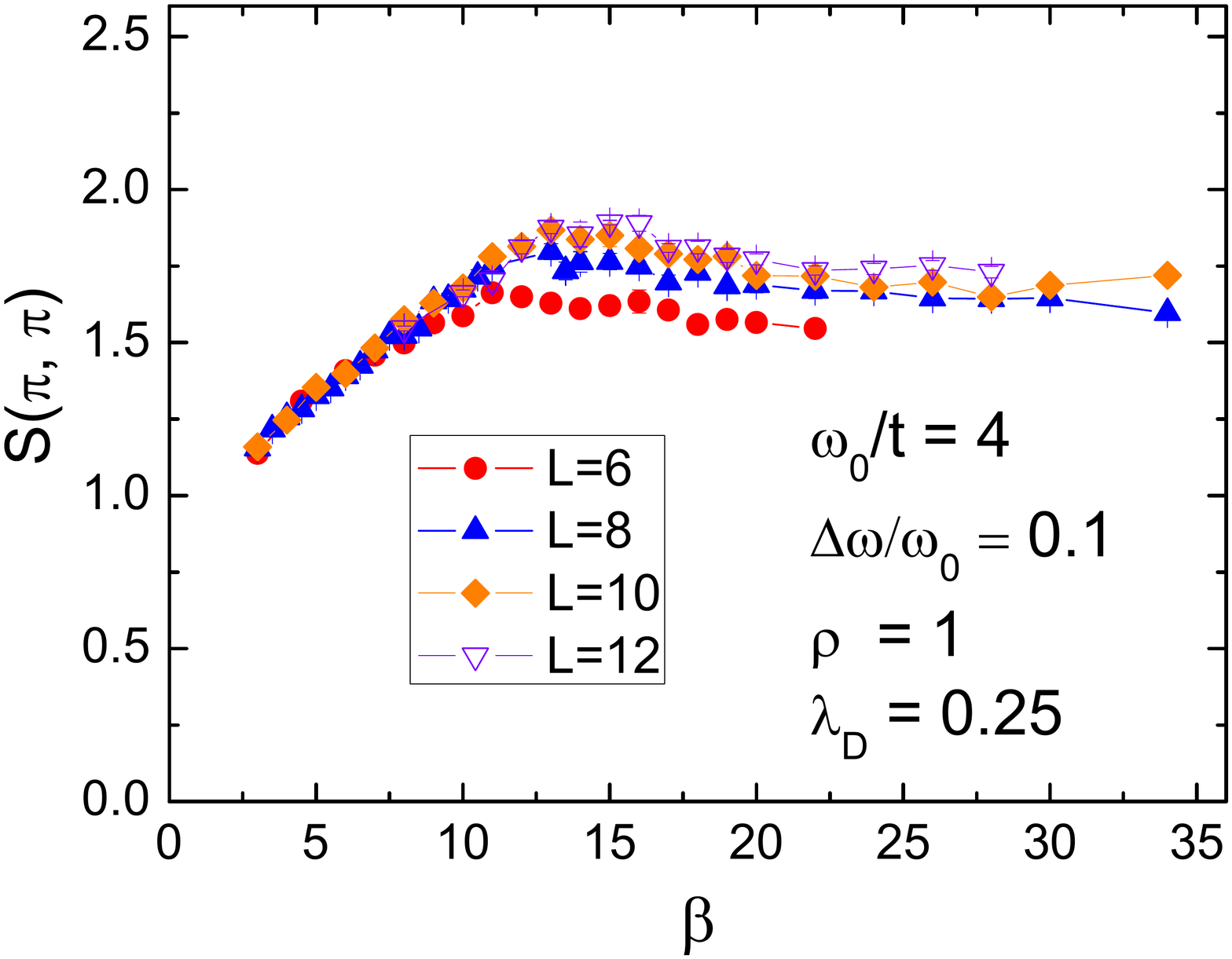} %
\caption{(Color online) Staggered chage-density structure factor behavior
at different lattice sizes for the same parameters of Fig.\,6.}
\label{fig:SF_wo40wa065}
\end{figure}

\vskip0.03in \noindent
\underbar{$S(\pi,\pi) / P_{s}$ at $\omega_{0}=4$ and $\Delta\omega=0$:}
For fixed $\lambda_{D}=0.25$ and $\omega_{0}/t=4.0$, the $S(\pi,\pi)$
of the dispersionless HM exhibits a strong dependence with lattice size
for $\beta \gtrsim 8$, as displayed in Fig.\,\ref{fig:wo40wa00}\,(a).
Performing the scaling [Eq.\,\eqref{eq:scaling}] of this raw DQMC data, as
presented in the inset, one finds $\beta_{c} \approx 13$.
This value of $T_{\mathrm{cdw}}(\omega_{0}=4)$ is lower than
$T_{\mathrm{cdw}}(\omega_{0}=1)$, even though we have kept $\lambda_{D}=0.25$
in both cases.
This happens because of the similarities between the Holstein model and the
attractive Hubbard model when $\omega_{0} \to \infty$.
The latter exhibits a degenerate ground state at half filling, with
coexistence between SC and CDW, called a supersolid
state\cite{Batrouni95}. In analogy to spin systems, this corresponds to
having both $xy$ (SC) and $z$ (CDW) components of the spin,
in which the Mermin-Wagner theorem forbids finite temperature transitions.
In view of this, one should expect
$T_{\mathrm{cdw}}(\omega_{0}) \to 0$ when $\omega_{0} \to \infty$.

On the other hand, the $s$-wave pair susceptibility exhibits very little
dependence with lattice size, as displayed
in Fig.\,\ref{fig:wo40wa00}\,(b), which
supports the picture of a ground state without SC.


\vskip0.03in \noindent
\underbar{$P_{s}$ at $\omega_{0}=4$ and $\Delta\omega \neq 0$:}
For completeness we present in Fig.\,\ref{fig:SF_wo40wa065} the
charge-density structure factor for the same parameters of Fig.\,6.
This result suggests that CDW is absent at $\Delta\omega/\omega_{0} = 0.10$.

\end{document}